\newcommand{\imagedir}{plots}

\documentclass[a4paper,11pt]{article}
\usepackage{jinst-tmp}
\def\havetwocolumns{0}

\usepackage{amsmath}
\usepackage{ifthen}
\usepackage{textcomp}
\usepackage{subfigure}

\title{A Layer Correlation technique for pion energy calibration at the 2004 ATLAS Combined Beam Test}

\author[k,1]{E.~Abat ,\note{Deceased}}
\author[f]{J.M.~Abdallah,}
\author[ag]{T.N.~Addy,}
\author[cc]{P.~Adragna,}
\author[ba]{M.~Aharrouche,}
\author[cm,2]{A.~Ahmad,\note{Now at SUNY, Stony Brook, United States of America}}
\author[ay]{T.P.A.~Akesson,}
\author[s]{M.~Aleksa,}
\author[n]{C.~Alexa,}
\author[t]{K.~Anderson,}
% ...

\emailAdd{kjg@particle.kth.se}

% ...
\affiliation[f]{Institut de Fisica d'Altes Energies, IFAE, Universitat Aut\`onoma de Barcelona, Edifici Cn, ES - 08193 Bellaterra (Barcelona) Spain}
% ...
\affiliation[k]{Bogazici University, Faculty of Sciences, Department of Physics, TR - 80815 Bebek-Istanbul, Turkey}
% ...
\affiliation[n]{National Institute of Physics and Nuclear Engineering (Bucharest -IFIN-HH),  P.O. Box MG-6, R-077125 Bucharest, Romania}
% ...
\affiliation[s]{European Laboratory for Particle Physics (CERN), CH-1211 Geneva 23, Switzerland}
\affiliation[t]{University of Chicago,  Enrico Fermi Institute,  5640 S. Ellis Avenue, Chicago, IL 60637, United States of America}
% ...
\affiliation[ag]{Hampton University, Department of Physics, Hampton, VA 23668, United States of America}
% ...
\affiliation[ay]{Lunds universitet, Naturvetenskapliga fakulteten,  Fysiska institutionen,  Box 118, SE - 221 00, Lund, Sweden}
% ...
\affiliation[ba]{Universit\"at Mainz,  Institut f\"ur Physik,  Staudinger Weg 7,  DE 55099,  Germany}
% ...
\affiliation[cc]{Queen Mary, University of London, Mile End Road, E1 4NS,  London, United Kingdom}
% ...
\affiliation[cm]{Insitute of Physics, Academia Sinica, TW - Taipei 11529, Taiwan}
% ...

\abstract{A new method for calibrating the hadron response of a
  segmented calorimeter is developed and successfully applied to beam
  test data. It is based on a principal component analysis of energy
  deposits in the calorimeter layers, exploiting longitudinal shower
  development information to improve the measured energy
  resolution. Corrections for invisible hadronic energy and energy
  lost in dead material in front of and between the calorimeters of
  the ATLAS experiment were calculated with simulated Geant4 Monte
  Carlo events and used to reconstruct the energy of pions impinging
  on the calorimeters during the 2004 Barrel Combined Beam Test at the
  CERN H8 area. For pion beams with energies between
  $20\ \mathrm{GeV}$ and $180\ \mathrm{GeV}$, the particle energy is
  reconstructed within 3\% and the energy resolution is improved by
  between 11\% and 25\% compared to the resolution at the
  electromagnetic scale.}


\begin{document}
\maketitle\pagestyle{myplain}\flushbottom

\section{Introduction}

In the general case of non-compensating calorimeters, the response to
hadrons will be lower than the response to particles which only
interact electromagnetically, such as electrons and photons. This is
due to energy lost in hadronic showers in forms not measurable as an
ionization signal, i.e., nuclear break-up, spallation and excitation,
energy deposits arriving out of the sensitive time window (such as
delayed photons), soft neutrons, and particles escaping the
detector~\cite{calorimetry1,calorimetry2,wigmanscalo}. Moreover, the
calorimeter response will be non-linear, since a hadronic shower has
both an electromagnetic and a hadronic component, with the size of the
former increasing with shower energy~\cite{groom}. In addition, the
large phase space of hadronic interactions leads to substantial
fluctuations in the size of the electromagnetic shower component from
event to event, degrading the measured energy resolution.

ATLAS~\cite{detectorpaper} is one of the multi-purpose physics
experiments at the CERN Large Hadron Collider
(LHC)~\cite{lhcpaper}. Scientific goals include searching for the
Higgs boson and looking for phenomena beyond the Standard Model of
particle physics, such as supersymmetry. Many measurements to be
performed by the LHC experiments rely on a correct and accurate energy
reconstruction of hadronic final-state particles. In the central
barrel region, the ATLAS calorimeters consist of the lead--liquid
argon (LAr) electromagnetic calorimeter
%~\cite{detectorpaper,lartdr}
and the Tile steel--scintillator hadronic calorimeter.
%~\cite{detectorpaper,tiletdr}.
Both calorimeters are intrinsically non-compensating.

Various techniques for equalizing the electromagnetic and hadronic
shower response, i.e., achieving compensation, have been proposed. For
a review, see~\cite{wigmanscalo}, chapter 3. Software-based offline
calibration techniques can use the topology of the visible deposited
energy to exploit spatial event-by-event information on shower
fluctuations and derive energy corrections aimed at restoring
linearity in the response and improving the energy resolution. For
example, the calorimeter cell energy density has been used for the
calorimeter in the H1 experiment~\cite{Issever2005803} and is planned
to be used in ATLAS~\cite{barillari:_local_hadron_calib}.

In this study, a calibration technique based on Monte Carlo simulation
is developed to deal with compensating the response of a segmented
calorimeter to hadrons and correcting for energy lost in the dead
material between two calorimeter systems. The correlations between
longitudinal energy deposits of the shower have been
shown~\cite{calocorr} to contain information on the electromagnetic
and hadronic nature of the shower. This information is utilized by
making a principal component analysis of the energies deposited in the
different calorimeter layers. The calibration is applied to pion beam
test data, taken at the 2004 ATLAS Barrel Combined Beam
Test~\cite{tancredinote,jenyanote,jenyapaper,electronpaper,ctb04}. The
method presented here is an alternative to the standard ATLAS
calibration schemes. The application is quite specific to ATLAS, but
the framework is general and it can be tested on any segmented
calorimeter. Energy corrections based on the longitudinal shower
development have been proposed by ATLAS in the context of jet
calibration~\cite{singleisolated900,singleisolated7,cscbooklayerweighting}.

The following section explains the basic principles of the
method. Section \ref{CBT2004} details the ATLAS Barrel Combined Beam
Test, while sections \ref{sec:emcalib} and \ref{evselpid} discuss
calibration to the electromagnetic scale and event selection,
respectively. The Geant4 Monte Carlo simulation used is described in
section \ref{mcsim}. Then, section \ref{implementation} gives the
details of the implementation of the calibration method. In section
\ref{sec:validmc}, the method is validated based on Monte Carlo
simulations of pions. In the Monte Carlo simulation, the effect of the
compensation weights and the dead material corrections are evaluated
separately. Lastly, the linearity and resolution of the final
calibrated energy is considered. Section \ref{sec:systematics}
discusses systematic uncertainties. Results of applying the method to
real beam test data are presented in section
\ref{subsec:perfdata}. Finally, conclusions are drawn in section
\ref{conclusions}.

\section{The Layer Correlation method}
\label{method}
The Layer Correlation calibration method (LC in the following) is
aimed at calibrating the response of a non-compensating longitudinally
segmented calorimeter to hadrons. Exploiting the properties of
hadronic showers to characterize fluctuations in the deposited
invisible energy, it uses a principal component analysis~\cite{PCA} of
the energy deposited in the calorimeter layers. Observables that
describe the shower fluctuations should be able to discriminate
between different corrections to be applied to recover invisible
losses due to hadronic interactions. Through the principal component
analysis, it is possible to reduce the number of dimensions that the
corrections depend on, while still capturing a large amount of event
fluctuation information and maintaining a good separation between
events with different content of invisible energy.

To derive the corrections, the interaction of the shower particles
with the detector material is simulated with the
Geant4~\cite{geant1,geant2} Monte Carlo simulation toolkit. In the
simulation the true energy deposited in the calorimeters and the
non-instrumented material is known. The covariance matrix between the
calorimeter layer energy deposits is calculated. Diagonalizing it, a
new orthogonal basis in the space of layer energy deposits is
derived. It consists of the eigenvectors of the covariance matrix. By
sorting the eigenvectors in descending eigenvalue order, the
projection of the energy deposits in the calorimeter layers along the
first few eigenvectors are made to describe the most important
fluctuations in the longitudinal shower development.

Using this information, compensation weights---correcting for the
non-compensation of the calorimeters---are derived in the form of
two-dimensional look-up tables in the projections along the first two
eigenvectors of the covariance matrix. One table is used for each
calorimeter layer. The tables are thus functions of two different
linear combinations of the observed energy deposits in the layers.

In addition, energy losses in non-instrumented material (so-called
``dead material'') will vary depending on the shower development. In
the ATLAS barrel region, these losses are primarily in the region
between the LAr and Tile calorimeters. The eigenvectors of the
covariance matrix considered above can also be used to correct for
this, resulting in a unified treatment for compensation and dead
material correction by deriving both corrections from the same set of
observables. In this implementation, the dead material corrections
have an inherent dependence on the beam energy. This dependence is
removed by employing an iteration scheme, where at each step the
estimated energy of the former step is used, until the returned value
is stable. A detailed mathematical description of the method is given
in section~\ref{implementation}.

\section{The 2004 ATLAS Barrel Combined Beam Test}
\label{CBT2004}
\begin{figure}
  \begin{center}
   \ifthenelse{\havetwocolumns=1}
   {\includegraphics[width=\linewidth]{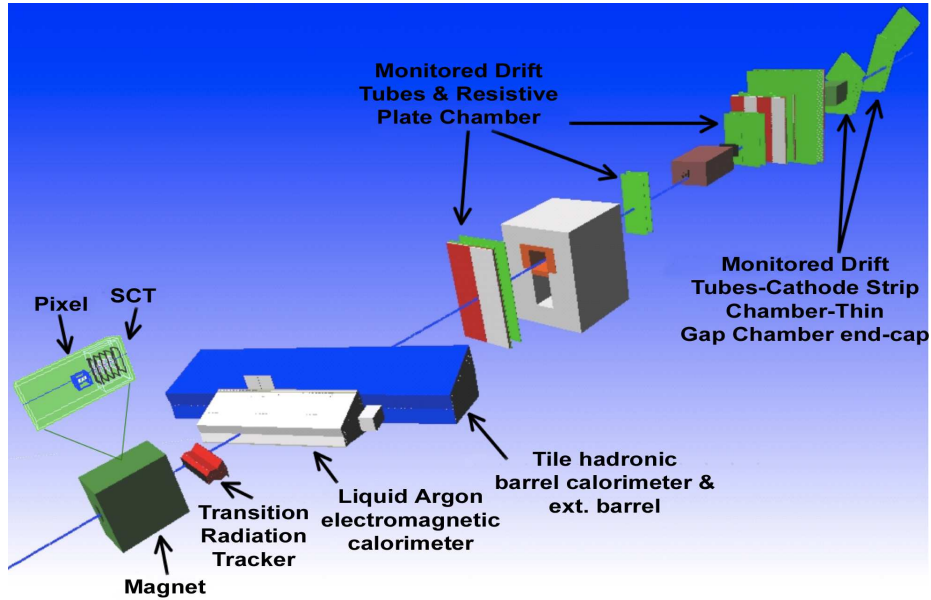}}
   {\includegraphics[width=0.8\linewidth]{\imagedir/bul-pho-2004-029.eps}}
   \end{center}
   \caption{The layout of the 2004 Combined Beam Test.}
   \label{fig:ctbdrawing}
\end{figure}

The energy calibration procedure is applied to data gathered in the
fall of 2004 during the ATLAS Barrel Combined Beam Test at the H8 beam
line of the CERN SPS accelerator. A full slice of the ATLAS barrel
region was installed (see Figure~\ref{fig:ctbdrawing}). This included,
firstly, the inner tracker with the pixel detector, the silicon strip
semiconductor tracker (SCT), and the straw tube transition radiation
tracker (TRT); secondly, the LAr and Tile calorimeters; and thirdly,
the muon spectrometer. The pixel and SCT detectors were surrounded by
a magnet capable of producing a field of $2\ \mathrm{T}$, although no
magnetic field was applied in the runs used for this study.

The pixel detector~\cite{detectorpaper} comprises six modules, each
consisting of a single silicon wafer with an array of $40 \times
400\ \hbox{\textmu}\mathrm{m^2}$ pixels. The modules were arranged in
locations mimicking the ATLAS configuration, with an approximate angle
of 20 degrees with respect to the incoming beam. The semiconductor
tracker (SCT)~\cite{detectorpaper} uses sets of stereo strips for
tracking. Each module gives two hits, one in each direction. Eight
modules, corresponding to those in the ATLAS end-cap, were used. The
TRT~\cite{detectorpaper} forms the outermost tracking system in
ATLAS. It consists of a collection of 4~mm diameter polyimide straw
tubes filled with a mixture of xenon, carbon dioxide, and
oxygen~\cite{detectorpaper}. Transition radiation is emitted when a
charged particle crosses the interface between two media having
different refractive index. The amount of emitted radiation depends on
the Lorentz $\gamma$ factor of the particle. This makes it possible to
discriminate between electrons and hadrons, given the much higher
$\gamma$ factor of the former at a given energy, due to their smaller
mass.

Details of the ATLAS LAr electromagnetic calorimeter are described
elsewhere~\cite{detectorpaper, lartdr}. In the beam test one
calorimeter module was used. The calorimeter is made from 2.21~mm
thick accordion-shaped lead absorbers glued between stainless steel
cathodes. Three-layered anode electrodes are interleaved between the
absorbers, spaced by 2~mm gaps over which a high voltage of 2~kV is
applied. The module was placed in a cryostat containing liquid
argon. The signal is read out by capacitive coupling between the two
outermost and the central layer of the anodes. In front of this
accordion module a thin presampler module was mounted. It consists of
two straight sectors with alternating cathode and anode electrodes
glued between plates made of a fiber-glass epoxy composite (FR4). The
Tile hadronic calorimeter consists of iron absorbers sandwiched
between organic scintillator tiles. It is described in detail
elsewhere~\cite{detectorpaper, tiletdr}. The tiles and absorbers are
oriented parallel to the direction of incoming particles. Every cell
of the calorimeter is read out by two wavelength-shifting fibers,
which in turn are grouped together and read out by photo-multiplier
tubes (PMTs).

\begin{figure*}%[htbp]
  \begin{center}
   \ifthenelse{\havetwocolumns=1}
   {\includegraphics[width=0.7\linewidth]{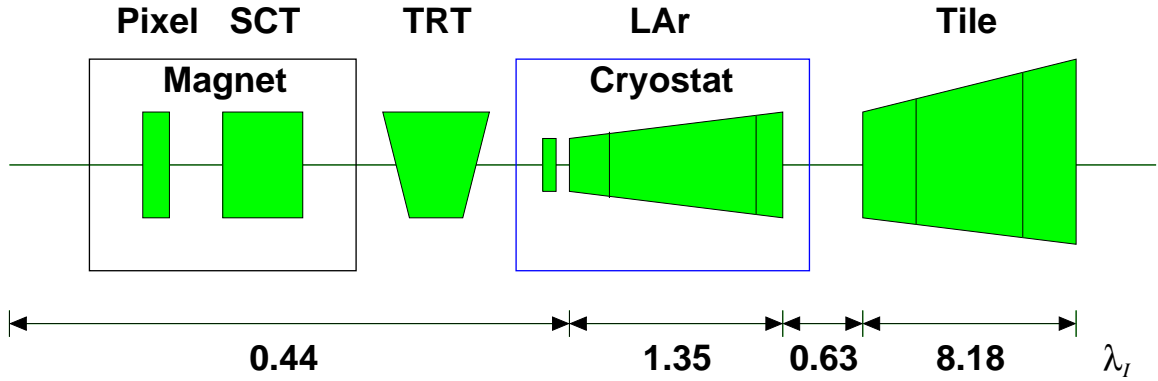}}
   {\includegraphics[width=\linewidth]{\imagedir/ctbkjg.eps}}
   \end{center}
   \caption{The layout of the 2004 Combined Beam Test.}
   \label{fig:ctblayout}
\end{figure*}

The calorimeters were placed so that the beam impact angle
corresponded to a pseudo-rapidity\footnote{ATLAS has a coordinate
  system centered on the interaction point, with the $x$ axis pointing
  towards the center of the LHC ring, the $y$ axis pointing straight
  up, and the $z$ axis parallel to the beam. Pseudo-rapidity is
  defined as $-\ln(\tan(\theta/2))$, where $\theta$ is the angle to
  the positive $z$ axis.} of $\eta$ = 0.45 in the ATLAS detector. At
this angle, the expected amount of material in front of the
calorimeters was about 0.44~$\lambda_{I}$, where $\lambda_{I}$ is the
nuclear interaction length~\cite{wigmanscalo, pdg}. This includes the
LAr presampler. The LAr calorimeter proper is longitudinally segmented
in three layers that extend in total for 1.35~$\lambda_{I}$. The dead
material between the LAr and Tile calorimeters spans about 0.63
$\lambda_{I}$. Finally the three longitudinal segments of the Tile
calorimeter stretched in total for about 8.18~$\lambda_{I}$. A sketch
of this setup is shown in Figure~\ref{fig:ctblayout}. In total there
are seven longitudinal calorimeter layers (the LAr presampler; the
front, middle, and back layers of the LAr calorimeter; and the
so-called A, BC, and D layers of the Tile calorimeter).  The length of
the individual calorimeter layers was 0.32, 0.96, and
0.07~$\lambda_{I}$ in the LAr calorimeter and 1.61, 4.53, and
2.04~$\lambda_{I}$ in the Tile calorimeter.

In addition, special beam-line detectors were installed to monitor the
beam position and reject background events. Those include beam
chambers monitoring the beam position and trigger scintillators. Beams
consisting of electrons, photons, pions, protons, and muons were
studied. In this analysis, pion beams with nominal momenta of 20, 50,
100, and 180~GeV were used (see Table~\ref{tab:datasample}). Data
belong to the fully combined run period, where all detector
sub-systems were present and operational. No magnetic field was
applied around the pixel and silicon strip detectors. The beams were
produced by letting 400~GeV protons from the SPS accelerator impinge
on a beryllium target, from which secondary pions are selected. For
the run at 180~GeV, positrons were nominally selected after the
target. However, the beam still contained a contamination of
positively charged pions, which were selected and used for this
analysis with the methods described in section~\ref{evsel}.

\section{Calorimeter calibration to the electromagnetic scale}
\label{sec:emcalib}
\subsection{Cell energy reconstruction}
\label{emcalib}
The individual cells of the calorimeter are calibrated to the
electromagnetic scale, i.e., with the aim of correctly measuring the
energy deposited in the cell by a purely electromagnetic shower. The
calibration of electronics of the LAr calorimeter is described in
detail in~\cite{ctblarcalib}. The method of optimal
filtering~\cite{cleland94:_signal} is used to reconstruct the
amplitude of the shaped signal, which is sampled by an ADC
(analog-to-digital converter) at $40\ \mathrm{MHz}$. The amplitude is
calculated as weighted sum of the samples, after a pedestal level
measured using random triggers is
subtracted. $F_{\mathrm{\hbox{\textmu}A \to MeV}} /
f_{\textrm{samp}}$, a constant factor, converts the measured current
to an energy measured in MeV. The energy deposited in the lead
absorbers is taken into account by the sampling fraction
$f_{\textrm{samp}}$. The shaping electronics is calibrated by
inserting calibration pulses of known amplitude. In the Tile
calorimeter a parameterized pulse shape is fitted to the samples. A
charge injection system is used to calibrate the read-out electronics,
while a cesium source is used to equalize the cell response, including
the response of the PMTs (see, for example~\cite{tilestandalone}).

\subsection{Topological clustering}
\label{topocluster}
Calorimeter cells calibrated to the electromagnetic scale are combined
by adding up the energy in neighboring cells using a topological
cluster algorithm~\cite{topocluster}. The algorithm has three
adjustable thresholds: Seed ($S$), Neighbor ($N$), and Boundary
($B$). First, seed cells having an energy above the $S$ threshold are
found and a cluster is formed starting with this cell. Then,
neighboring cells having an energy above the $N$ threshold are added
to the cluster. This process is repeated until the cluster has no
neighbors with an energy above the $N$ threshold. Finally, all
neighboring cells having an energy above the $B$ threshold are added
to the cluster. To avoid bias, the absolute values of the cell
energies are used. The $S$, $N$, and $B$ thresholds are set to,
respectively, four, two, and zero times the expected noise standard
deviation in the cell considered.

\subsection{Pion energy reconstruction}
The reconstructed energy in a calorimeter layer $L$ is obtained by
considering all the topological clusters in the event and summing up
the parts of the clusters that are part of that calorimeter layer. The
total reconstructed energy is then derived by summing over the
$N_\textrm{lay}$ longitudinal layers in the calorimeter.

\section{Event selection and particle identification}
\label{evselpid}

\subsection{Event selection}
\label{evsel}
A signal in the trigger scintillator and a measurement in adjacent
beam chambers that is compatible with one particle passing close to
the nominal beam line are required. In addition, exactly one track,
where the sum of the number of hits in the Pixel detector and the SCT
is more than six, is asked for, as well as at least 20 hits in the
TRT. The track in the TRT must be compatible with a pion track, i.e.,
no more than two hits passing the high threshold must be
present. Events with a second track in the TRT are rejected: this
ensures that the pion does not interact strongly before the
TRT. Furthermore, there must be at least one topological cluster (see
section \ref{topocluster}) with at least 5 GeV in the calorimeter.
This cut rejects muons contained in the beam and does not influence
the pion energy measurement. To reject some residual electron
background, events with more than 99\% of their energy in the LAr
calorimeter are excluded. The same selection is applied on simulated
Monte Carlo events as on data, with the exception of cuts related to
the beam chambers and scintillators.

\subsection{Proton contamination}
\label{protoncontamination}
This study used beams of pions with positive electric charge. These
beams are known to have a sizable proton contamination
$f_{\textrm{prot}}$ defined as the fraction of events in a sample that
result from protons impinging on the calorimeters. It varies between
different beam energies. The TRT makes it possible to measure the
average proton contamination of the test beam for each beam energy,
owing to the different probabilities between pions and protons of
emitting transition radiation, although it is not possible to
discriminate between the particles on an event-by-event basis. The
measured~\cite{tancredinote} contamination is reported in
Table~\ref{tab:datasample}. For the 20~GeV beam energy, a one-sided
confidence interval is given. In the analysis, a proton contamination
of 0\% was used. Agreement is found with measurements performed by a
\v Cerenkov counter at a 2002 beam test~\cite{tilecalmargar} conducted
in the same beam line.

\begin{table*}
  \begin{center}
    \begin{tabular}{|c|c|c|c|c|}\hline
      $E_\textrm{beam}^\textrm{nom}$ (GeV) & $E_\textrm{meas}$ (GeV)&
      No. ev. bef. cuts &
      No. ev. after cuts &
      $f_\textrm{prot}$\\ 
      \hline \hline
      \phantom{1}20  & \phantom{1}20.16 & \phantom{1}49871 & \phantom{2}8957  &  $<$ 17\% (84\% CL)\\
      \phantom{1}50  & \phantom{1}50.29 &           109198 &           29578  &  (\phantom{-}45 $\pm$ 12)\%\\
      100            & \phantom{1}99.89 & \phantom{1}67220 & \phantom{1}5843  &  (\phantom{-}61 $\pm$ \phantom{1}6)\%\\ 
      180            & 179.68           &           105082 &           11780  &  (\phantom{-}76 $\pm$ \phantom{1}4)\%\\ \hline
    \end{tabular}
  \end{center}
    \caption{Data samples taken in the 2004 Combined Beam Test used in the present analysis.}
      \label{tab:datasample}
\end{table*}

\section{Monte Carlo simulation}
\label{mcsim}
\subsection{Hadronic shower simulation}
All calibration corrections are extracted from a Geant4.7
\cite{geant1,geant2} Monte Carlo simulation, with an accurate
description of the Combined Beam Test geometry. The physics list---
i.e., set of models---QGSP\_BERT was used. It uses the
QGSP~\cite{qgsp} (Quark Gluon String Pre-compound) phenomenological
model describing the hadron--nucleus interaction by the formation and
fragmentation of excited strings together with the de-excitation of an
excited nucleus. The Bertini
model~\cite{qgspbertini1,qgspbertini2,qgspbertini3} of the
intra-nuclear hadronic cascade is used to describe nuclear
interactions at low energies. This model treats the particles in the
cascade as classical and propagates them through the nucleus, which is
modeled as a medium with a density averaged in concentric
spheres. Excited states are collected and the nucleus decays in a
slower phase following the fast intra-nuclear cascade.

The Bertini model is applied up to an energy of 9.9~GeV, while the
QGSP model applies from 12 GeV and upward. In an intermediate range of
9.5--25~GeV, the low-energy parameterized LEP
model~\cite{geant_physic_refer_manual} is used. In the energy ranges
where models overlap, the decision which one to use is made
stochastically using a continuous linear probability distribution that
goes from exclusively using the low-energy model at the lower end of
the region to exclusively using the high-energy model at the upper
end.

\subsection{Detector simulation}

The simulation provides not only reconstructed calorimeter cell
energies at the electromagnetic scale---including the effects of the
readout electronics---but also the true deposited energy, which is
divided into four components: electromagnetic visible, hadronic
visible, invisible, and escaped. Visible energy results from
ionization of the calorimeter material. Invisible energy is energy not
directly measurable in the detector, such as break-up energy in
nuclear interactions. The escaped energy represents the small
contribution from neutrinos, high-energy muons and, possibly, neutrons
and low-energy photons escaping the total simulated volume.

\subsection{Event samples}
\label{evsamples}
Monte Carlo samples were produced by simulating both pions and protons
impinging on the detector setup. Two statistically independent event
samples were produced by dividing the available sample into two
approximately equal parts: one set (``correction'' samples in the
following) was used to derive compensation weights and dead material
corrections, while the other set (``signal'' samples in the following)
was used to validate the weighting procedure and find the expected
performance. Pions and protons were simulated at 25 different beam
energies, ranging from $15\ \mathrm{GeV}$ to $230\ \mathrm{GeV}$. In
total, about 800 000 events per sample and particle type were
available after event selection. The energy spacing was $2$, $3$, or
$5\ \mathrm{GeV}$ up to $70\ \mathrm{GeV}$ and 10 or
$20\ \mathrm{GeV}$ above $70\ \mathrm{GeV}$. This spacing was found to
give satisfactory performance (see sections \ref{sec:validmc} and
\ref{subsec:perfdata}). Further studies of different spacings can be
pursued when applying this technique to different calorimeters to
explore possible improvement in performance.

Taking the proton beam contamination mentioned in section
\ref{protoncontamination} into account, all the available
``correction'' Monte Carlo samples were used to build a ``mixed''
pion--proton sample, one for each energy available in the data (see
Table \ref{tab:datasample}). Each of these samples is used as input
when deriving the corrections used for that proton fraction. In this
way the corrections were tuned to the studied proton fraction. If the
samples had different numbers of events, a sample-dependent weight was
first applied to give them equal weight before selection cuts.  Then,
given the proton contamination $f_\textrm{prot}$ at a given energy,
pion and proton events for each same-energy pair of samples were
assigned a weight of $1 -f_\textrm{prot}$ and $f_\textrm{prot}$,
respectively.

\section{Implementation of the Layer Correlation method}
\label{implementation}
\subsection{Calculation of the eigenvectors of the covariance matrix}
\label{subsec:covmatrixcalc}

Each event is associated with a set of $N_\textrm{lay}$ layer energy
deposits ($E^\textrm{rec}_{1}$,
...\ ,\ $E^\textrm{rec}_{N_\textrm{lay}}$), one per calorimeter layer,
representing a point in an $N_\textrm{lay}$-dimensional vector space,
referred to in the following as the space of layer energy
deposits. They are reconstructed energies at the electromagnetic
scale, formed as calorimeter layer sums of topological clusters as
described in section~\ref{emcalib}. The $N_\textrm{lay}$-dimensional
covariance matrix of the layer energy deposits is calculated as
\begin{align}
\textrm{Cov}(M,L) = \left \langle{}E_{M}^\textrm{rec}E_{L}^\textrm{rec} \right\rangle{} -
\left\langle{}E_{M}^\textrm{rec}\right\rangle{}\left \langle{}E_{L}^\textrm{rec}\right\rangle{},
\label{eq:covmatrix}
\end{align}
where $M$ and $L$ denote calorimeter layers and $E_{M}^\textrm{rec}$
is the energy reconstructed at the electromagnetic scale in
calorimeter layer $M$. The averages are defined as
\begin{align}
\langle{}E_{M}^\textrm{rec}E_{L}^\textrm{rec}\rangle{}\ = \frac{\sum_{i} E_{M,i}^\textrm{rec}
E_{L,i}^\textrm{rec}}{N_{\mathrm{ev}}} \quad \textrm{and} \quad \langle{}E_{M}^\textrm{rec}\rangle{}\ = \frac{\sum_{i} E_{M,i}^\textrm{rec}}{N_{\mathrm{ev}}}.
\label{eq:defcovmatrix}
\end{align}
The sums are performed over all the $N_{\mathrm{ev}}$ events in the
sample. The eigenvectors of the covariance matrix form a new
orthogonal basis in the space of layer energy deposits. The
coordinates of the point in the $N_\textrm{lay}$-dimensional vector
space corresponding to an event $i$ can be expressed in this new
eigenvector basis as
\begin{align}
 E_{\textrm{eig},M}^\textrm{rec} = \sum_L
 \alpha^\textrm{rec}_{M,L} E^\textrm{rec}_L,
\label{eq:coordinates}
\end{align}
where $\alpha^\textrm{rec}_{M,L}$ are the coefficients of the
transition matrix to the new basis. Projections of events along the
covariance matrix eigenvectors represent independent fluctuations. The
variances of those fluctuations are given by the corresponding
eigenvalues. The eigenvectors are sorted in descending order according
to their eigenvalues, meaning that the first eigenvectors determine
the directions along which most of the event fluctuations take
place. The layer energy covariance matrix Cov($M,L$)
(equations~\ref{eq:covmatrix} and~\ref{eq:defcovmatrix}) is calculated
using events from the ``mixed'' sample.

In any given event a symmetric energy cut is applied on each layer
energy such that the energy for that layer is re-defined as
$E_{L}^\textrm{rec}$, if $|E_{L}^\textrm{rec}|$ $\rangle{}$
$E_{L}^\textrm{thr}$, zero otherwise. The goal of such cuts is to
eliminate the contribution of noise-dominated layers. The energy
threshold values for each calorimeter layer can be found in
Table~\ref{tab:thresholds}. The cuts were optimized to obtain the best
expected compensation performance on Monte Carlo samples at 50 GeV.

\begin{table}
  \begin{center}
    \begin{tabular}{|c|c|}\hline
      Calorimeter layer & Threshold (GeV) \\
      \hline \hline
      0 & 0.032 \\
      1 & 0.108 \\
      2 & 0.030 \\
      3 & 0.150 \\
      4 & 0.039 \\
      5 & 0.070 \\
      6 & 0.042 \\ \hline
    \end{tabular}
  \end{center}
    \caption{Energy thresholds per calorimeter layer.}
      \label{tab:thresholds}
\end{table}

\begin{figure}
  \begin{center}
  \includegraphics[width=0.45\linewidth]{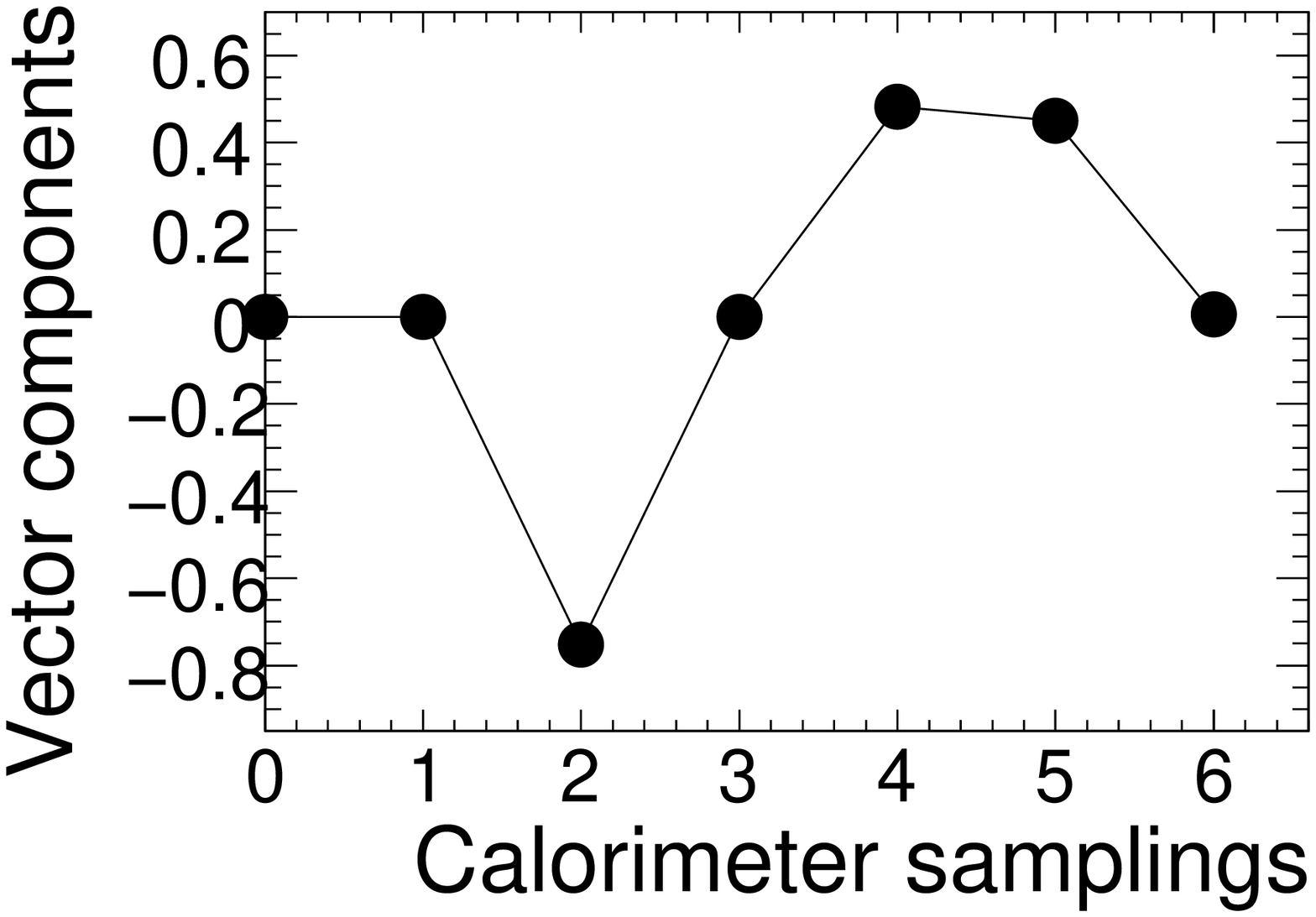}
  \includegraphics[width=0.45\linewidth]{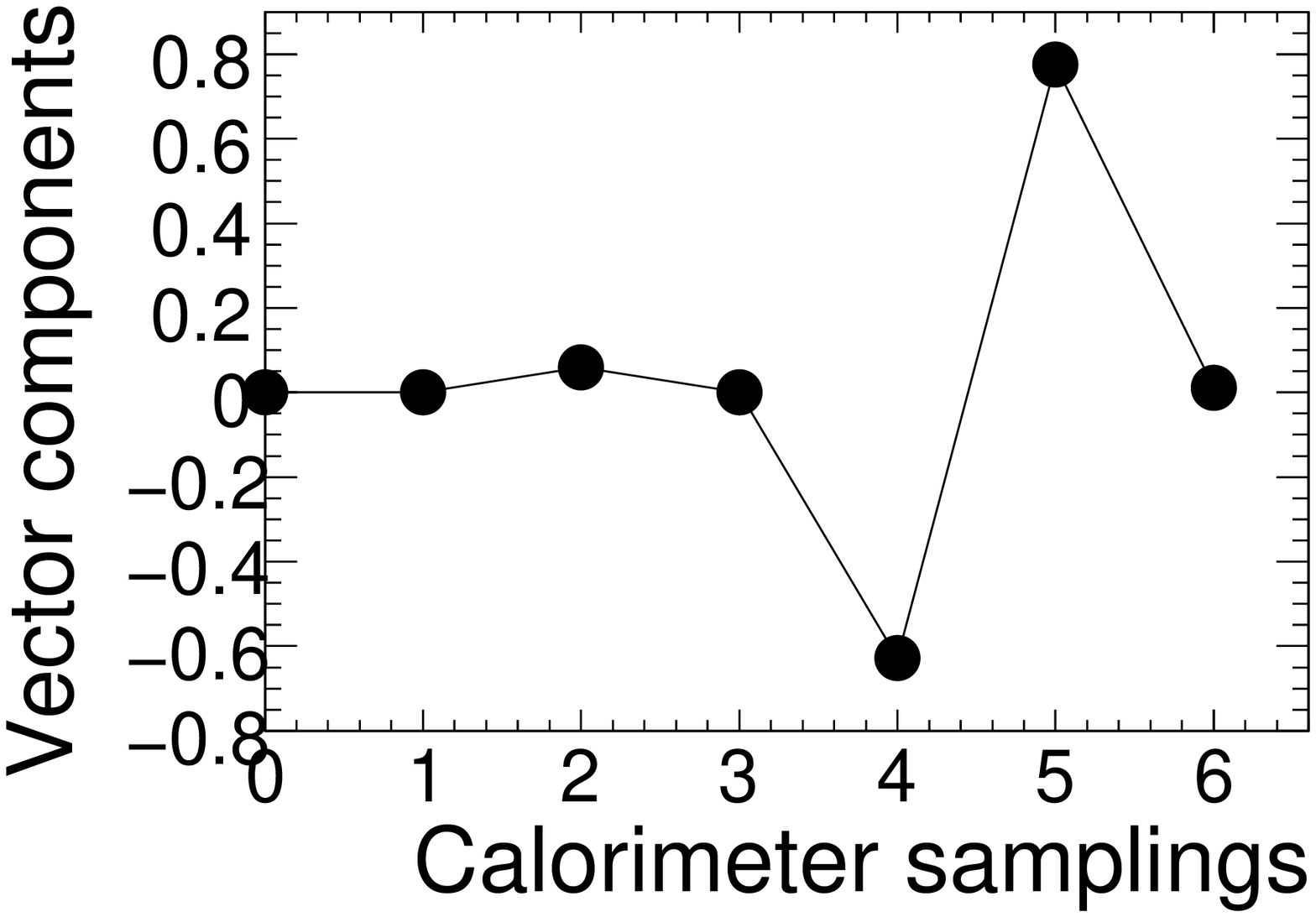}
  \includegraphics[width=0.45\linewidth]{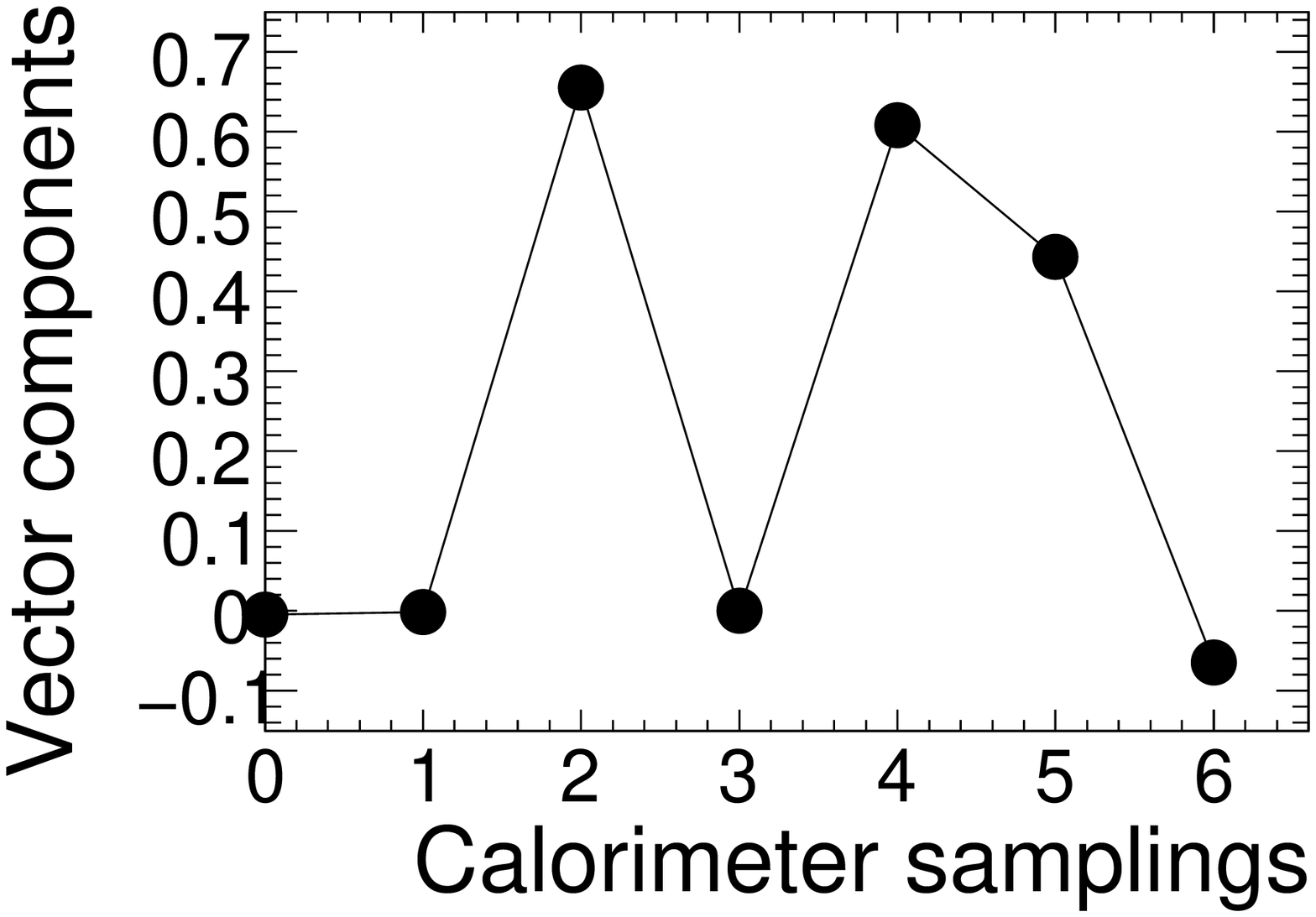}
  \end{center} \caption{Eigenvector components for the first three
  eigenvectors expressed in the basis of the seven layers of the ATLAS
  calorimeters in the Combined Beam Test for a simulated mix of
  protons and pions with 45\% proton
  contamination.\label{fig:eigenvecsmix}}
\end{figure}

A physical interpretation of the eigenvalues and normalized
eigenvectors can be obtained from Figure~\ref{fig:eigenvecsmix}, which
shows the components of the first three eigenvectors expressed in the
original basis of calorimeter layer energy deposits. We find that
\begin{align}
E^\textrm{rec}_{\textrm{eig},0} &\approx \frac{1}{\sqrt{6}} (-2 E_\textrm{LAr,middle} + E_\textrm{Tile,A} + E_\textrm{Tile,BC}),\label{eq:eig0}\\
E^\textrm{rec}_{\textrm{eig},1} &\approx \frac{1}{\sqrt{2}} (-E_\textrm{Tile,A}  + E_\textrm{Tile,BC}),\ \textrm{and}\label{eq:eig1}\\
E^\textrm{rec}_{\textrm{eig},2} &\approx \frac{1}{\sqrt{3}} (E_\textrm{LAr,middle} + E_\textrm{Tile,A} + E_\textrm{Tile,BC}).\label{eq:eig2}
\end{align}
So in a qualitative but suggestive way, we can make the interpretation
that $E^\textrm{rec}_{\textrm{eig},0}$ corresponds to the difference
between the Tile and LAr calorimeters, since most of the energy
deposited in the LAr calorimeter is deposited in the middle
layer. $E^\textrm{rec}_{\textrm{eig},1}$ corresponds to the difference
between the second and first layers of the Tile calorimeter, while
$E^\textrm{rec}_{\textrm{eig},2}$ corresponds to most of the energy of
the event. The other eigenvectors represent individual calorimeter
layers. These layers are rather thin and appear to be uncorrelated to
the other layers.

\subsection{Compensation weights}
\label{subsec:weighting}

The compensation weights account for the non-linear response of the
calorimeters to hadrons. There is one weight table for each
calorimeter layer, i.e., three for the LAr calorimeter and three for
the Tile calorimeter. The seventh layer, the LAr presampler, which in
order is the first layer, is not used in the weighting procedure, as
explained below. The total reconstructed energy is the sum of the
weighted energies in each calorimeter layer:
\begin{align}
E_{L}^\textrm{weighted} &= w_{L}  E_{L}^\textrm{rec} \\
E_\textrm{tot}^\textrm{weighted} &=  \sum_{L} E_{L}^\textrm{weighted}.
\end{align}

For each event $i$, there is an ideal set of $N_\textrm{lay}$
coefficients that would re-weight each reconstructed energy deposit in
layer $L$ to the true deposited energy:
\begin{align}
w^\textrm{ideal}_{L,i} = E^\textrm{true}_{L,i}/E^\textrm{rec}_{L,i}.
\label{eq:theidealweight}
\end{align}
The symbol $E_{L,i}^\textrm{rec}$ ($E_{L,i}^\textrm{true}$) denotes
the reconstructed (true) energy deposited in the $L^\textrm{th}$ layer
in the $i^\textrm{th}$ event. The task is to find a set of weights
$w_{L}$ that approximate the ideal weights. In general, for each layer
$L$, the weight is an $N_\textrm{lay}$-dimensional function of the
layer energy deposits. Exploiting the fluctuation-capturing properties
of the eigenvector projections, the weights can in general be derived
as a function of an $N$-dimensional subspace of the
$N_\textrm{lay}$-dimensional space of layer energy deposits, spanned
by the first $N$ eigenvectors. In the absence of an analytic
formulation, the layer weights $w_{L}$ are estimated by Monte Carlo
sampling: multi-dimensional cells are built, which partition the
$N$-dimensional vector space along the directions of the base
eigenvectors. In general, these cells are multi-dimensional
hyper-cubes. They are referred to as bins below.

For each bin $k$ one defines the weight as the average of the ideal
weights of equation \ref{eq:theidealweight}:
\begin{align}
w_{k,L} =\ \langle{} E^\textrm{true}_{L,i}/E^\textrm{rec}_{L,i} \rangle{}_k\ = \frac{1}{N_{\mathrm{ev},k}} \sum_{i} E^\textrm{true}_{L,i}/E^\textrm{rec}_{L,i},
\label{eq:avweight}
\end{align}
where the summation is performed for the $N_{\mathrm{ev},k}$ events in
the bin. If each event has a weight\footnote{For instance, to equalize
 the number of events for all data sets.} $p_i$, the average is
modified accordingly:
\begin{align}
w_{k,L} =\ \langle{} E^\textrm{true}_{L,i}/E^\textrm{rec}_{L,i} \rangle{}_k\ = \frac{\sum_{i} p_i E^\textrm{true}_{L,i}/E^\textrm{rec}_{L,i}}{\sum_{i} p_i}.
\label{eq:avweight_we}
\end{align}
Using bin $k$ of the weight tables, the total reconstructed energy
becomes
\begin{align}
E_{\textrm{tot},k}^\textrm{weighted} &= \sum_{L} w_{k,L}
E_{L}^\textrm{rec}.
\end{align}
Here, the $w_{k,L}$ functions defined in equation~\ref{eq:avweight_we}
are estimated in bins of the two-dimensional space spanned by the
eigenvectors corresponding to the two highest eigenvalues, i.e.,\ $N =
2$. Thus each layer is associated with a two-dimensional look-up
table. For a given layer the average weights in each two-dimensional
bin are calculated using only the energy values that passed the cuts
defined in section~\ref{subsec:covmatrixcalc}. The table has the same
number of equally spaced bins along the two dimensions: $128 \times
128$. Bi-linear interpolation is performed between the bins. Weights
for the LAr presampler are not calculated, even if the presampler is
kept in the covariance matrix. No weights are applied to the energy
deposited in the presampler layer, and energy deposited in the
presampler itself is taken as part of the upstream dead material
losses.

In addition the compensation weights and corrections derived from the
proton sample are corrected by the factor
\begin{align}
\frac{E_\textrm{beam}^\textrm{nom}}{E_\textrm{beam}^\textrm{nom} - m_\textrm{proton}},
\end{align}
where $m_\textrm{proton}$ is the proton mass, to account for the fact
that, for a proton, the sum of the total true deposited energy in the
calorimeter is $E_\textrm{beam}^\textrm{nom} - m_\textrm{proton}$.

\begin{figure}%[htbp]
  \begin{center}
    \ifthenelse{\havetwocolumns=1}
    {\subfigure[]{\label{fig:weighttables:subfig:a}\includegraphics[width=\columnwidth,angle=0]{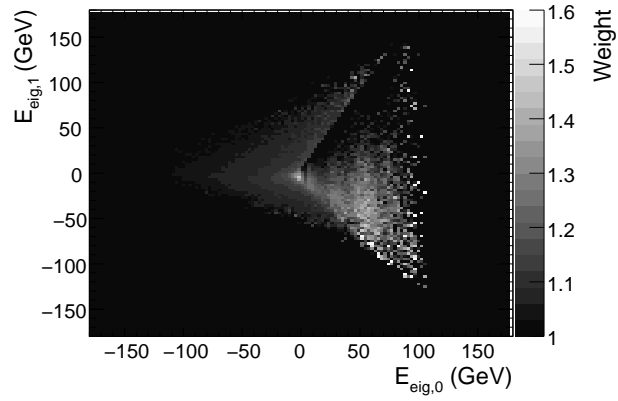}}
     \subfigure[]{\label{fig:weighttables:subfig:b}\includegraphics[width=\columnwidth,angle=0]{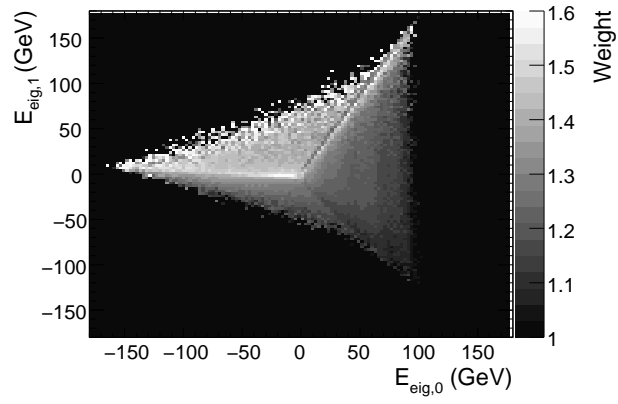}}
     \subfigure[]{\label{fig:weighttables:subfig:c}\includegraphics[width=\columnwidth,angle=0]{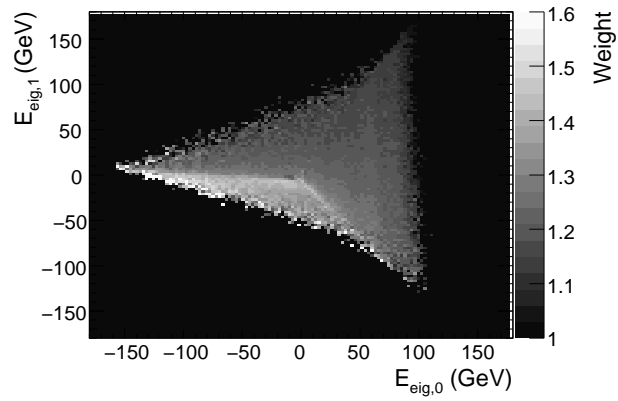}}}
    {\subfigure[]{\label{fig:weighttables:subfig:a}\includegraphics[width=0.55\columnwidth,angle=0]{\imagedir/weighttable_2.eps}}
     \subfigure[]{\label{fig:weighttables:subfig:b}\includegraphics[width=0.55\columnwidth,angle=0]{\imagedir/weighttable_4.eps}}
     \subfigure[]{\label{fig:weighttables:subfig:c}\includegraphics[width=0.55\columnwidth,angle=0]{\imagedir/weighttable_5.eps}}}
    \end{center} \caption{Compensation weights as a function of the
    first two eigenvector projections for simulated pion--proton mixed
    events (45\% proton contamination) in the second layer of the LAr
    calorimeter (a), first layer of the Tile calorimeter (b), and
    second layer of the Tile calorimeter (c).\label{fig:weighttables}}
\end{figure}

Typical compensation weight tables are shown in Figure
\ref{fig:weighttables}: they illustrate the look-up tables for the
second (middle) layer of the LAr calorimeter and for the first and
second layer of the Tile calorimeter for a pion--proton mixed sample
with 45\% contamination. The triangular shape visible in the weight
tables can be understood from the interpretation of the eigenvectors
of equations \ref{eq:eig0} and \ref{eq:eig1}. With increasing energy
in the Tile calorimeter and less in the LAr calorimeter, i.e.,
$E^\textrm{rec}_{\textrm{eig},0}$ is large, there are more values that
can be assumed by $E^\textrm{rec}_{\textrm{eig},1}$, which is the
approximate difference between the first and second layers of the Tile
calorimeter. Three lines can be seen extending from the origin to each
of the three corners of the triangle. Firstly, the line extending from
the origin and to the left corresponds to events where close to all of
the energy is deposited in the LAr calorimeter. The small slope is due
to the slight dependence of $E^\textrm{rec}_{\textrm{eig},1}$ on the
second layer of the LAr calorimeter. Secondly, the line extending up
and to the right corresponds to events where all energy is deposited
in the second layer of the Tile calorimeter. Along that line, weights
are small for the first sampling of the Tile calorimeter, since
particles are still minimum-ionizing in that layer. Thirdly the faint
line extending down and to the right corresponds to events where close
to all the energy is deposited in the first layer of the Tile
calorimeter.

\subsection{Dead material corrections}
\label{subsec:dmcorr}

Regions of dead material constitute those parts of the experiment that
are neither active calorimeter read-out material (liquid argon or
scintillator), nor sampling calorimeter absorbers (mostly lead or
steel). The LC technique is used for the dead material between the LAr
and the Tile calorimeters, while a simple parameterized model is
utilized for other losses.

\subsubsection{Dead material between the LAr and Tile calorimeters}
\label{subsubsec:dmlartile}
Most of the dead material is in the LAr cryostat wall between the LAr
and Tile calorimeters. In this 0.6$~\lambda_I$ region, pion showers
are often fully developed, giving rise to large energy loss. Each
event $i$ is associated with a point in the layer energy deposit
vector space as explained in section \ref{subsec:covmatrixcalc}. It
also has a true total energy lost in the dead material between the LAr
and Tile calorimeters: $E^\textrm{DM,true}_\textrm{LArTile}(i)$. The
dead material correction $E^\textrm{DM}_{\textrm{LArTile}}$ for each
event $i$ can be derived as a $T$-dimensional function of the layer
energy deposits. In general, the subspace chosen for deriving the dead
material correction and its dimension $T$ can be different from the
one chosen for compensation, both in content (spanned by different
eigenvectors) and in dimension ($T$ can be different from $N$). The
value of $E^\textrm{DM}_{\textrm{LArTile}}$ is estimated by Monte
Carlo sampling. For any $T$-dimensional bin $m$ one defines
\begin{align}
E^\textrm{DM}_{\textrm{LArTile},m} =\ \langle{}E^\textrm{DM,true}_{\textrm{LArTile},i}\rangle{}_m,
\label{eq:avdm}
\end{align}
where the average is performed for the events in that bin.

Here, the correction defined in equation~\ref{eq:avdm} is calculated
in bins of the two-dimensional space spanned by the eigenvectors
corresponding to the first and third eigenvalues, i.e.,\ $T = 2$. This
was the combination of eigenvectors that was found to give the best
performance. As for the compensation weights, correction tables are
derived from a $128 \times 128$ bin look-up table and bi-linear
interpolation is performed between the bins.

\begin{figure}%[htbp]
  \begin{center}
   \ifthenelse{\havetwocolumns=1}
   {\includegraphics[width=\columnwidth,angle=0]{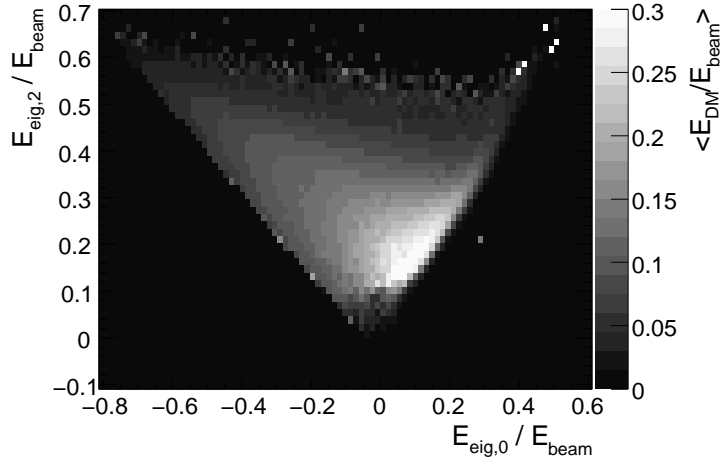}}
   {\includegraphics[width=0.64\columnwidth,angle=0]{\imagedir/dmtable.eps}}
   \end{center} \caption{Look-up table for LAr--Tile dead material
   corrections as a function of the first and third eigenvector
   projections normalized to beam energy for 45\% proton
   contamination.~\label{fig:dmtablewholescan}}
\end{figure}

The three dimensions of the look-up table are all shown to scale with
the beam energy, i.e., a table determined at a given beam energy can
be turned into one at a different beam energy by scaling all the
dimensions with the ratio of the two energies. Consequently, all
dimensions in the table---the eigenvector projections and the average
dead material losses---are divided by the beam energy when filling the
table. That is, the event coordinates in the space of layer energy
deposits are expressed as
\begin{align}
 E_{\textrm{eig},M}^\textrm{rec,norm}= E_{\textrm{eig},M}^\textrm{rec}/E = \sum_L \alpha^\textrm{rec}_{\textrm{eig},L} E^\textrm{rec}_L/E,
\label{eq:coordinatesdm}
\end{align}
where the variables have the same meaning as in
equation~\ref{eq:coordinates} and $E$ is the best estimate of the beam
energy of the simulated pion in that event (see below). The dead
material look-up table is shown in Figure~\ref{fig:dmtablewholescan}
for a pion--proton mixed sample with 45\% contamination. The figure
shows the distribution of the rescaled dead material energy as a
function of the rescaled event coordinates.  Regions with different
dead material fractions can be differentiated. They range between 0
and more than 30\% of beam energy. In addition, the samples at
different energies behave very similarly as a function of the
re-scaled variables.

\subsubsection{Other dead material corrections}
\label{subsubsec:otherdmcorr}

\begin{figure}
  \begin{center}
    \ifthenelse{\havetwocolumns=1}
    {\includegraphics[width=\columnwidth,angle=0]{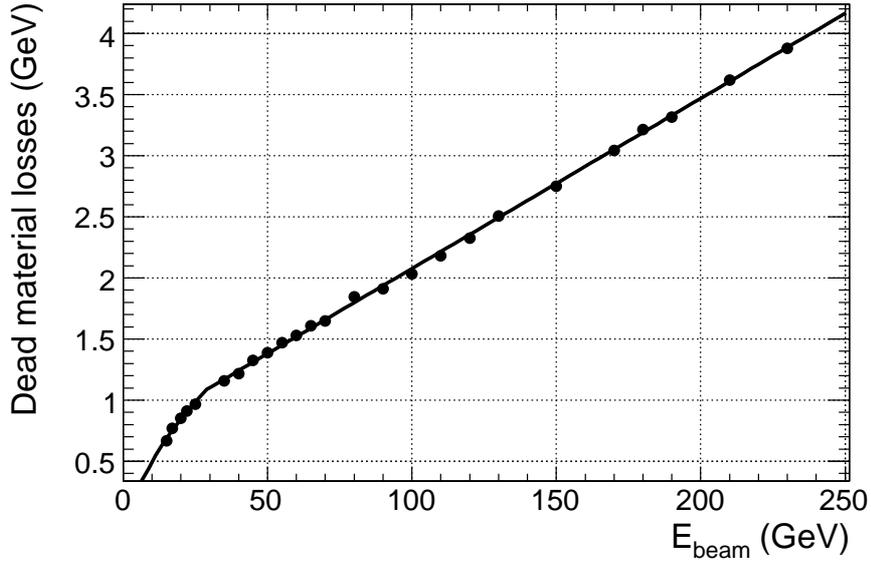}}
    {\includegraphics[width=0.8\columnwidth,angle=0]{\imagedir/dm_other.eps}}
    \end{center} \caption{Mean dead material losses other than those
    between the LAr and Tile calorimeters as a function of the beam
    energy. Filled circles indicate the mean loss obtained from Monte
    Carlo simulation. The line indicates a parameterization to
    interpolate between the beam energies.~\label{fig:leakagefit}}
\end{figure}

While the energy losses between the LAr and Tile calorimeters
dominate, there are still other regions where dead material losses can
occur. These are losses located in the material upstream of the LAr
calorimeter, between the LAr presampler and the first LAr calorimeter
layer, and energy leakage beyond the Tile calorimeter. To compensate
for these losses the mean energy loss was determined as a function of
beam energy and the resulting data points were fitted using a suitable
functional form
\begin{align}
E^\textrm{DM}_\textrm{other}(E_\textrm{beam}) = \left\{
 \begin{array}{rl}
  C_1 + C_2\,\sqrt{E_\textrm{beam}} & \quad \textrm{if}\quad E_\textrm{beam} < E_0 \\ 
  C_3 + C_4\,(E_\textrm{beam} - E_0) & \quad \textrm{otherwise,}
 \end{array}\right.
\label{eq:otherdmcor}
\end{align}
where $E_0 = 30\ \mathrm{GeV}$. As an example, the fit for a proton
fraction of 45\% can be seen in Figure \ref{fig:leakagefit}. The
resulting fitted parameters are
\begin{align}
C_1 & = (-353 \pm 23)\ \mathrm{MeV},\\
C_2 & = (8.47 \pm 0.17)\ \sqrt{\mathrm{MeV}},\\
C_3 & = (1102 \pm 3)\ \mathrm{MeV},\quad \textrm{and}\\
C_4 & = 0.01392 \pm 0.0001.
\end{align}

\subsection{Applying the calibration}
\label{subsec:finalcalib}

The final energy after calibration consists of the sum of the weighted
calorimeter layer energies and the dead material corrections:
\begin{align}
E_{\textrm{tot}, k, m}^\textrm{corr} &= E_{\textrm{tot},k}^\textrm{weighted} + E^\textrm{DM}_{\textrm{tot},m}(E).
\label{formula:weighteden}
\end{align}
The index $k$ stands for the bin in the appropriate $N$-dimensional
space of layer energy deposits used in the weight tables (equations
\ref{eq:avweight} or \ref{eq:avweight_we}), while $m$ is the bin in
the $T$-dimensional space of layer energy deposits used to build the
LC estimate for the energy loss in the dead material between the LAr
and Tile calorimeters obtained from equation~\ref{eq:avdm}. The total
dead material correction is derived from summing the two contributions
derived in sections~\ref{subsubsec:dmlartile}
and~\ref{subsubsec:otherdmcorr}:
\begin{align}
E^\textrm{DM}_{\textrm{tot},m}(E)=E^\textrm{DM}_{\textrm{LArTile},m}+E^\textrm{DM}_\textrm{other}(E),
\label{eq:totaldm}
\end{align}
where $E$ is the best estimate for the total deposited pion energy
used to estimate $E_\textrm{beam}$ in equation~\ref{eq:otherdmcor}.

The events in a Monte Carlo sample are usually generated at a fixed
beam energy in order to test the calorimeter response. Corrections
derived from a fixed beam energy sample are, in principle, dependent
on that information, i.e., they depend on the same quantity (pion
energy) for the reconstruction of which they should be used. For the
compensation weights, this dependence is overcome by superposing
events from all the available energies. The eigenvector projections
scale approximately with the energy of the incoming particle, meaning
that regions in the table that come in use for a certain particle
energy will be dominated by samples close to that energy.

On the other hand, the look-up-table-based LAr--Tile dead material
correction and the parameterized model for the other dead material
losses have an inherent dependence on an assumed beam energy when
applying the corrections (see equations \ref{eq:coordinatesdm} and
\ref{eq:totaldm}). This dependence is overcome using an iteration
technique, giving the end result of depending only on the energy in
the calorimeters. At each step the best estimate of the reconstructed
energy $E_\textrm{tot}^\textrm{corr}$ after all corrections is used to
set both the scaling factor $1/E$ (equation~\ref{eq:coordinatesdm})
for the LAr--Tile correction and the best pion energy estimate in the
parameterization for the other dead material corrections. Each new
estimate of the energy is used to pick up a new correction from the
look-up table until the returned value is stable. In the initial step
$E_\textrm{tot}^\textrm{corr}$ is just the pion energy after
compensation weights are applied. The iteration cut-off is a tunable
parameter.

The process of applying the calibration is as follows:
\begin{itemize}
\item Associate each event to a bin in both the $N$-dimensional
  compensation weight and the $T$-dimensional dead material correction
  spaces defined in sections~\ref{subsec:weighting}
  and~\ref{subsec:dmcorr} by expressing its electromagnetic-scale
  energy deposit vector in the new eigenvector basis derived from the
  simulated events.
\item Extract compensation corrections for the energy of each given
  layer and the LAr--Tile dead material correction from the look-up
  tables. Apply all corrections according to
  equations~\ref{formula:weighteden} and~\ref{eq:totaldm}.
\item Use the iteration for dead material corrections.
\end{itemize}

\section{Method validation on Monte Carlo simulation}
\label{sec:validmc}
Before applying it to beam test data, the calibration is validated on
a Monte Carlo sample statistically independent of the one used for
extracting the corrections. First, the performance of the compensation
weights is evaluated, then the linearity and resolution of the method
as a whole. The weighting technique is validated on Monte Carlo
simulation samples in separate steps:
\begin{itemize}
\item Reconstruct the true deposited energy in the calorimeters
  (compensation validation).
\item Reconstruct the full energy of the incoming particles, including
  dead material corrections, and quantify the performance in terms of
  linearity and resolution.
\end{itemize}
The performance is evaluated in terms of bias and resolution.  The
weights and dead material corrections are derived from the
``correction samples'' and applied on the statistically independent
``signal samples'' (see section \ref{evsamples}). The results in this
section are derived for pions only.

\subsection{Compensation validation}
\label{subsec:compensation}
The reconstructed pion energy after compensation correction is
compared to the true deposited energy in the calorimeter. The
event-by-event difference $E_\textrm{tot}^\textrm{weighted} -
E_\textrm{tot}^\textrm{true}(\textrm{calo})$ is considered, where
$E_\textrm{tot}^\textrm{true}(\textrm{calo})$ is the true total energy
deposited in the calorimeter. The bias in the energy reconstruction is
defined as the average value
$\langle{}E_\textrm{tot}^\textrm{weighted} -
E_\textrm{tot}^\textrm{true}(\textrm{calo})\rangle{}$ and the
resolution is obtained by calculating the standard deviation
$\sigma$($E_\textrm{tot}^\textrm{weighted} -
E_\textrm{tot}^\textrm{true}(\textrm{calo})$).

The performance of the LC technique is compared with a simple
calibration scheme (called $f_\textrm{comp}$ in the following) which
uses beam energy information: each event in the sample is weighted
with the same factor $f_\textrm{comp}$ =
$\langle{}E^\textrm{true}_\textrm{tot}\rangle{}$/$\langle{}E^\textrm{reco}_\textrm{tot}\rangle{}$,
where $\langle{}E^\textrm{true}_\textrm{tot}\rangle{}$
($\langle{}E^\textrm{reco}_\textrm{tot}\rangle{}$) is the average true
total (reconstructed) energy deposited in the given sample in the
whole calorimeter, but not in the dead material. The $f_\textrm{comp}$
calibration scheme provides a reference scale to which the improvement
in resolution of the LC weighting can be compared.

\begin{figure}
  \begin{center}
    \ifthenelse{\havetwocolumns=1}
    {\subfigure[]{\label{fig:biasweighting:subfig:bias}\includegraphics[width=\columnwidth,angle=0]{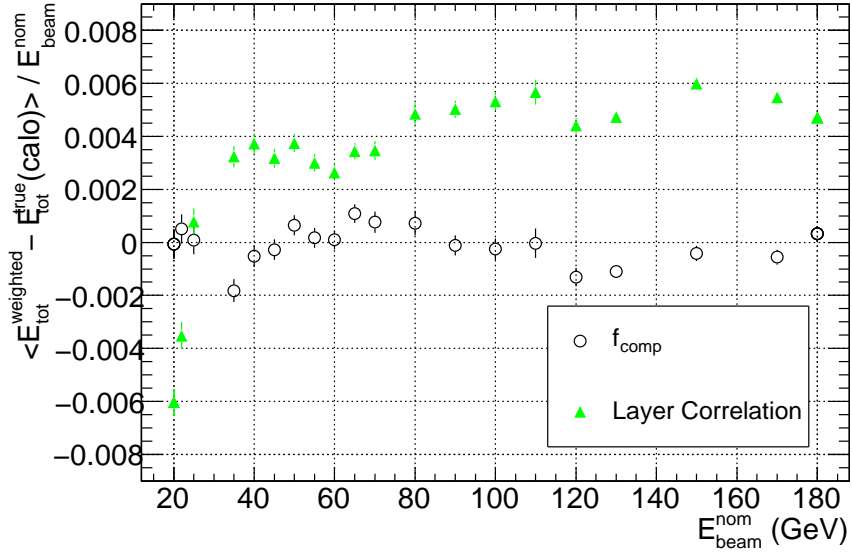}}
    \subfigure[]{\label{fig:biasweighting:subfig:res}\includegraphics[width=\columnwidth,angle=0]{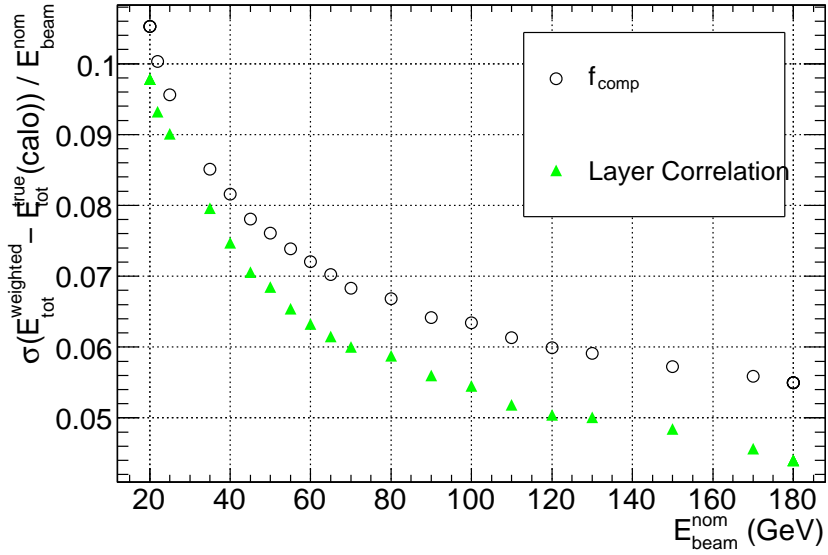}}}
    {\subfigure[]{\label{fig:biasweighting:subfig:bias}\includegraphics[width=0.8\columnwidth,angle=0]{\imagedir/mc_eoverpi_diff_mean.eps}}
    \subfigure[]{\label{fig:biasweighting:subfig:res}\includegraphics[width=0.8\columnwidth,angle=0]{\imagedir/mc_eoverpi_diff_rms.eps}}}
    \end{center} \caption{Bias (a) and resolution (b) of the
    reconstructed energy after compensation correction minus the true
    deposited energy for energy deposited in the calorimeters in
    simulated samples for the calibration procedure using a simple
    factor and LC weighting.~\label{fig:biasweighting}}
\end{figure}

The results of the validation procedure are shown in
Figure~\ref{fig:biasweighting}. By construction, there is no bias in
the energy reconstruction for the calibration procedure using a simple
factor. The LC weighting mostly gives a slight positive bias of about
0.6\%. At the lower edge of the energy range studied, the bias instead
turns slightly negative. The resolution improvement increases with
beam energy. It is about 10\% at 50~GeV and about 20\% at 180~GeV.

\subsection{Dead material corrections}

\begin{figure}[htbp]
  \begin{center}
    \ifthenelse{\havetwocolumns=1}
    {\includegraphics[width=\columnwidth,angle=0]{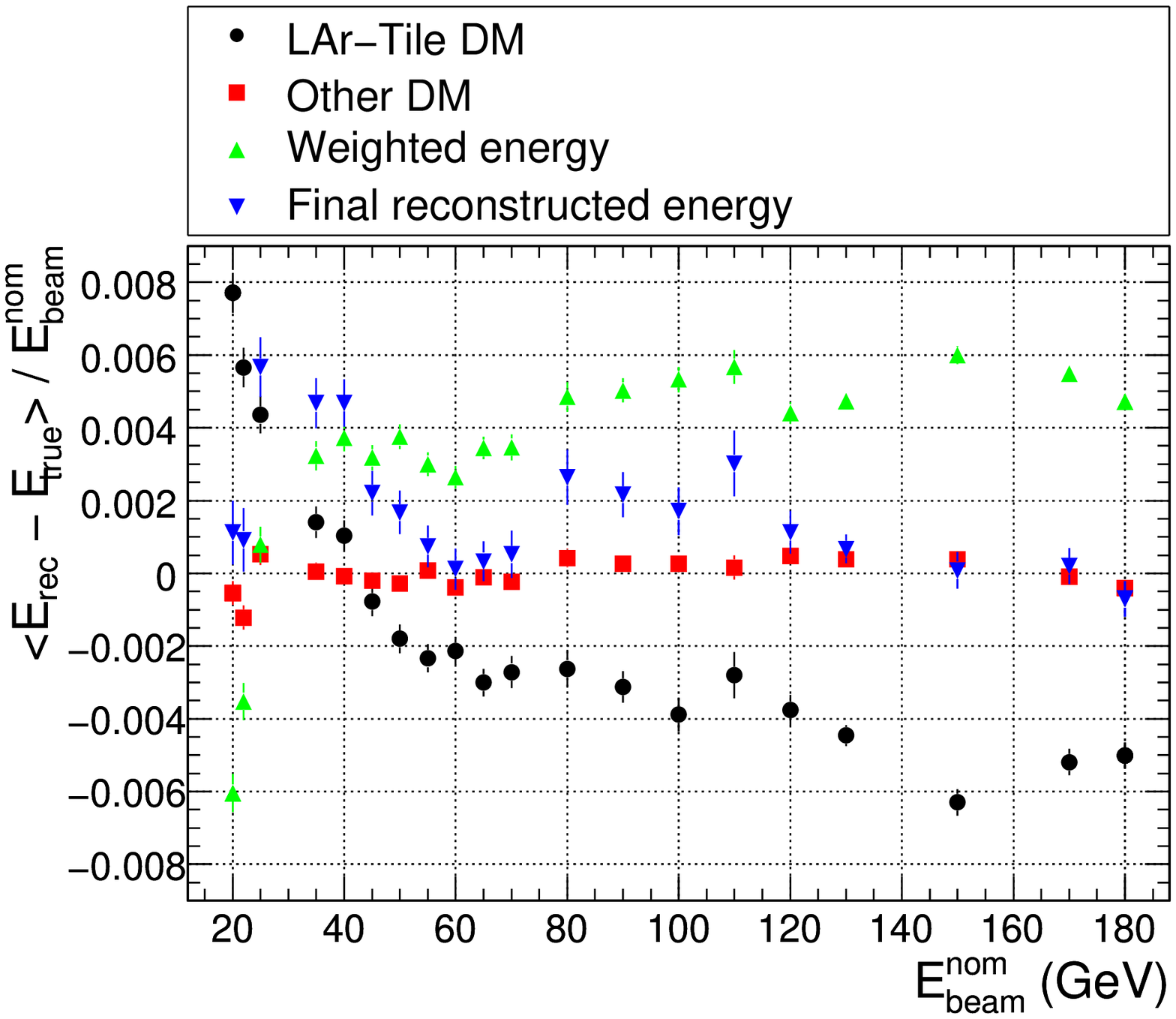}}
    {\includegraphics[width=0.75\columnwidth,angle=0]{\imagedir/mc_weighted_diff_mean.eps}}
    \end{center} \caption{Bias (reconstructed energy minus true
    deposited energy, divided by beam energy) for the three individual
    corrections: Weighted calorimeter energy, correction for dead
    material between the LAr and Tile calorimeters, and other
    dead material corrections. Lastly, the bias of the final
    reconstructed energy, which is the sum of the
    three.~\label{fig:bias}}
\end{figure}

Figure~\ref{fig:bias} shows the bias of the weighted energy, and also
the bias of the dead material corrections. For most energies, the
LAr--Tile dead material correction has a slight negative bias, while
at low energies the bias is positive. The bias is 0.5\%
maximally. This cancels out most of the bias from the weighting. The
final energy is reconstructed correctly within a few per mil.

\subsection{Linearity and resolution in the Monte Carlo sample}
\label{sec:resandlin}

The performance for the fully corrected energy reconstruction is
finally assessed in terms of linearity with respect to the beam energy
and relative resolution. The reconstructed energy distribution is
fitted with a Gaussian distribution in the interval ($\mu$ -
2$\sigma$, $\mu$ + 2$\sigma$), where $\mu$ and $\sigma$ are the mean
value and the standard deviation, respectively. This interval is found
iteratively. The mean value $E_\textrm{fit}$ and the standard
deviation $\sigma_\textrm{fit}$ of the fitted Gaussian are used
together with the beam energy $E_\textrm{beam}$ to define the
linearity and the relative resolution.
\begin{itemize}
\item The linearity is $E_\textrm{fit}$/${E_\textrm{beam}}$ as a function of $E_\textrm{beam}$.
\item The relative resolution is $\sigma_\textrm{fit}$/$E_\textrm{fit}$ as a function of $E_\textrm{beam}$. 
\end{itemize}
Both linearity and relative resolution are derived for the energy
distribution at four stages of the energy reconstruction:
\begin{itemize}
\item at the electromagnetic scale,
\item after applying the compensation weights,
\item after compensation weights and application of dead material
  correction for losses between the LAr and Tile calorimeters, and
\item after compensation weighting and all dead material corrections.
This last step aims to reconstruct the pion energy.
\end{itemize}

\begin{figure}
  \begin{center}
   \ifthenelse{\havetwocolumns=1}
   {\subfigure[]{\label{subfig:lin}\includegraphics[width=\columnwidth,angle=0]{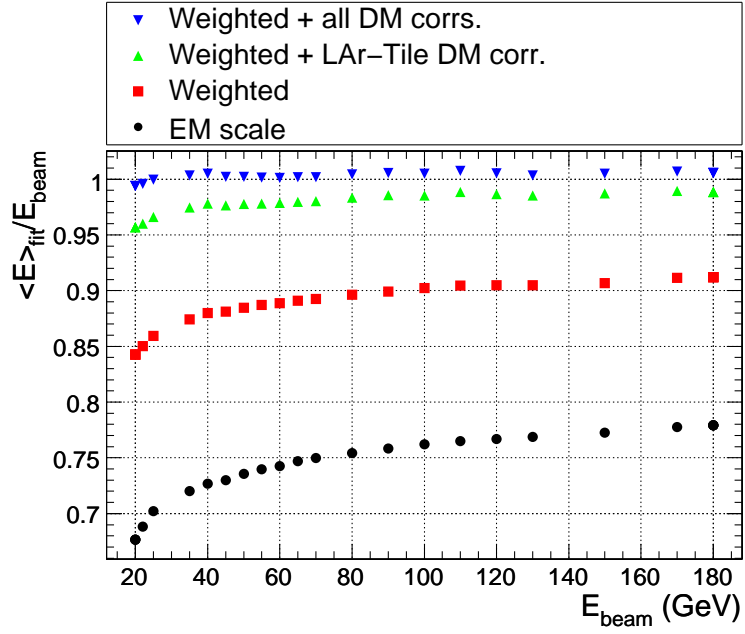}}
    \subfigure[]{\label{subfig:res}\includegraphics[width=\columnwidth,angle=0]{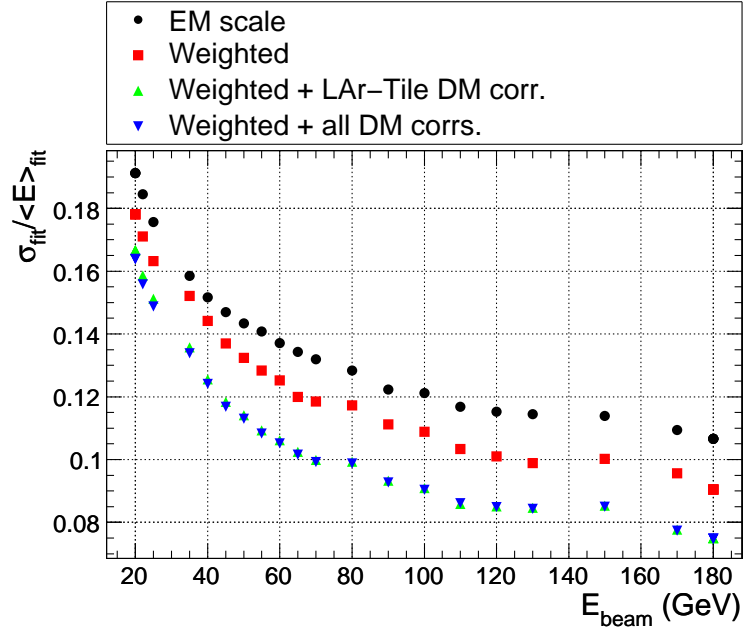}}}
   {\subfigure[]{\label{subfig:lin}\includegraphics[width=0.70\columnwidth,angle=0]{\imagedir/mc_lin.eps}}
    \subfigure[]{\label{subfig:res}\includegraphics[width=0.70\columnwidth,angle=0]{\imagedir/mc_res.eps}}}

   \end{center} \caption{Linearity (a) and relative resolution (b) for
   simulated pure pion samples at the electromagnetic scale, with
   compensation weights applied, compensation weights plus LAr--Tile
   dead material correction applied, and all corrections
   applied.~\label{fig:mcpionslinres_wholescan}}
\end{figure}

This is shown in Figure~\ref{fig:mcpionslinres_wholescan}. At the
electromagnetic scale the calorimeter response is non-linear---as
expected---and only about two thirds of the pion energy is
measured. After weighting, between 80\% and 90\% of the incoming pion
energy is recovered, while the dead material between the LAr and Tile
calorimeters accounts for an additional 8--10\%. After all corrections
the correct pion energy is reconstructed within 1\% for all beam
energies. Each correction step makes the calorimeter response more
linear. The compensation weights give a better improvement of the
linearity at high energies, while the dead material effects play a
more significant role at low energies, in particular at 20 GeV where
other corrections than LAr--Tile dead material are important to get to
within 1\% of the beam energy. The relative resolution is improved
when applying each of the different correction steps.\footnote{The
  apparent discontinuity in resolution between the results at energies
  below 150 GeV and those above might be due to a geometry change in
  the description of the beam test setup: three centimeters of
  aluminum were included in the Inner Detector system for energies
  larger than or equal to 150 GeV.}  At high beam energies (above
$E_\textrm{beam}$ = $100\ \textrm{GeV}$) the contribution of the
compensation weights to the improvement in energy resolution has the
same magnitude as that of the LAr--Tile dead material corrections.  At
lower beam energies dead material corrections account for about 70\%
of the relative resolution improvement down to about
$E_\textrm{beam}\simeq 30\ \textrm{GeV}$.  Below $E_\textrm{beam}$
$\simeq$ 30 GeV all the corrections account for a similar fraction of
the improvement: other dead material corrections than those for
LAr--Tile account for about 20\% of the resolution improvement, they
are marginal above that threshold.

\section{Systematic uncertainties}
\label{sec:systematics}
Systematic uncertainties on the calibrated energy can be divided into
\begin{itemize}
\item The uncertainty of the beam energy: 0.7\%~\cite{electronpaper}.
\item The absolute electromagnetic scale uncertainty, which is
  estimated to be 0.7\%~\cite{electronpaper} in the LAr calorimeter
  and 1.0\%~\cite{tilestandalone} in the Tile calorimeter. Scaling the
  cell energies with their corresponding uncertainties gives a
  combined electromagnetic scale uncertainty of 0.9\%.
\item The sensitivity of the results to the proton fraction at each
  beam energy. It was estimated by varying the fraction used to
  calculate the corrections. With the assumed fraction adjusted up or
  down one standard deviation of the TRT measurement, the relative
  variation in linearity and resolution in data and Monte Carlo simulation was
  found to be of the order of 1\% for $E_\textrm{beam} = 20\ \textrm{GeV}$
  and 50 GeV and less than 0.5\% for 100~GeV and 180~GeV.
\end{itemize}
Adding these contributions in quadrature gives a total systematic
uncertainty of less than 2\% for each beam energy.

\section{Application of the method to beam test data}
\label{subsec:perfdata}
Finally, the method is applied to beam test data, which is compared
with Monte Carlo samples with a weighted mixture of pions and protons
to match the beam composition.

\subsection{Data to Monte Carlo simulation comparison}
\begin{figure}
  \begin{center}
    \ifthenelse{\havetwocolumns=1}
    {\includegraphics[width=\columnwidth,angle=0]{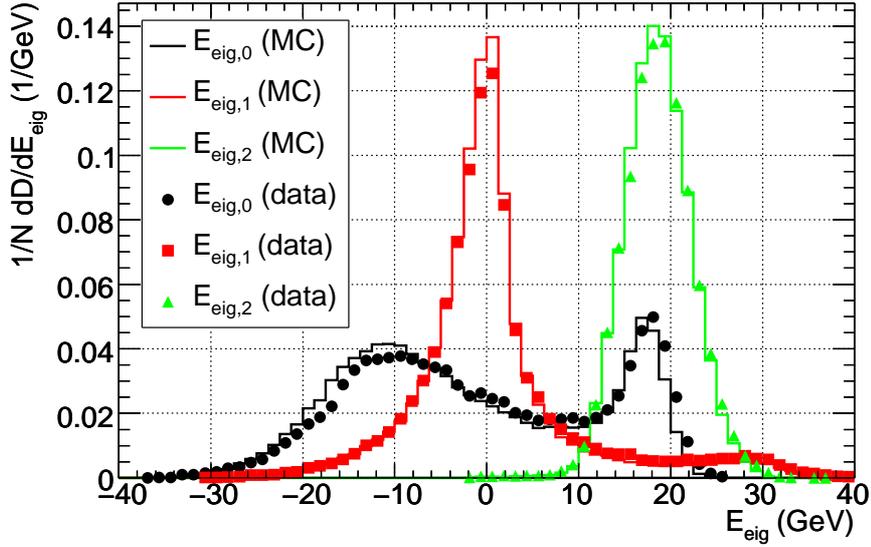}}
    {\includegraphics[width=0.8\columnwidth,angle=0]{\imagedir/eigvecprojs_50.eps}}
  \end{center} \caption{Distribution of the first three eigenvector
  components for data (filled circles) and Monte Carlo simulation pion--proton
  ``mixed signal'' with a proton fraction of 45\% and a beam energy of
  50~GeV.~\label{fig:eigenvec_mcvsdata}}
\end{figure}

The pion--proton ``mixed signal'' samples are used to compare data and
Monte Carlo simulations in terms of the distribution of the first
three components of the layer energy vector along the basis of
covariance matrix eigenvectors as defined in
section~\ref{subsec:covmatrixcalc}.
Figure~\ref{fig:eigenvec_mcvsdata} shows such a comparison for a
proton fraction of 45\% and a beam energy of 50~GeV. Good agreement is
obtained between data and simulation. The distribution for
$E_{\textrm{eig},0}$ shows a double peak structure that separates
events mainly showering in the Tile calorimeter from those where the
shower starts earlier.

\begin{figure}
  \begin{center}
    \ifthenelse{\havetwocolumns=1}
    {\includegraphics[width=\columnwidth,angle=0]{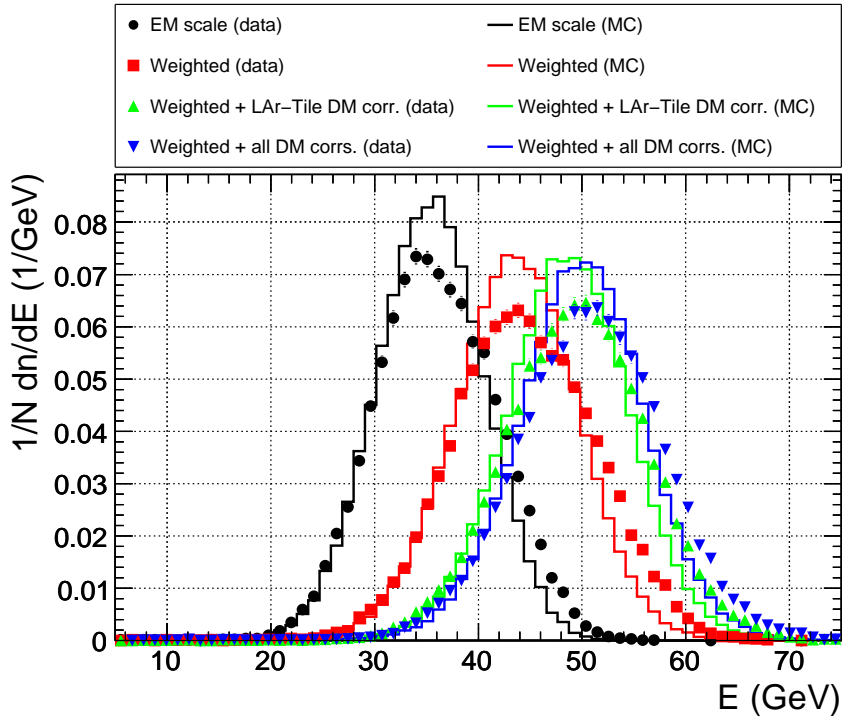}}
    {\includegraphics[width=0.8\columnwidth,angle=0]{\imagedir/hist_50_GeV.eps}}
    \end{center} \caption{Normalized energy distribution for
    $E_\textrm{beam}$ = 50 GeV after applying subsequent corrections
    for compensation and dead material effects. For the Monte Carlo
    simulation a proton fraction of 45\% is
    used.~\label{fig:compshape50}.}
\end{figure}

The shapes of the energy distributions (in unit bins of energy and
events) for data and Monte Carlo simulation are compared in
Figure~\ref{fig:compshape50}. The corrections are successively
applied. Already at the electromagnetic scale the energy distribution
is not well reproduced. The distribution in the Monte Carlo simulation
is narrower and less skewed than in the data. This effect is even
larger at 20~GeV but less pronounced at higher energies. The quality
of the initial description of data by Monte Carlo simulation is not
modified by the application of the compensation weights and dead
material corrections (see also section~\ref{sec:linresfinal}).

\subsection{Linearity and resolution on data}
\label{sec:linresfinal}
\begin{figure}
  \begin{center}
   \ifthenelse{\havetwocolumns=1}
   {\subfigure[]{\label{fig:linearity_data_mc_wholescan}\includegraphics[width=\columnwidth,angle=0]{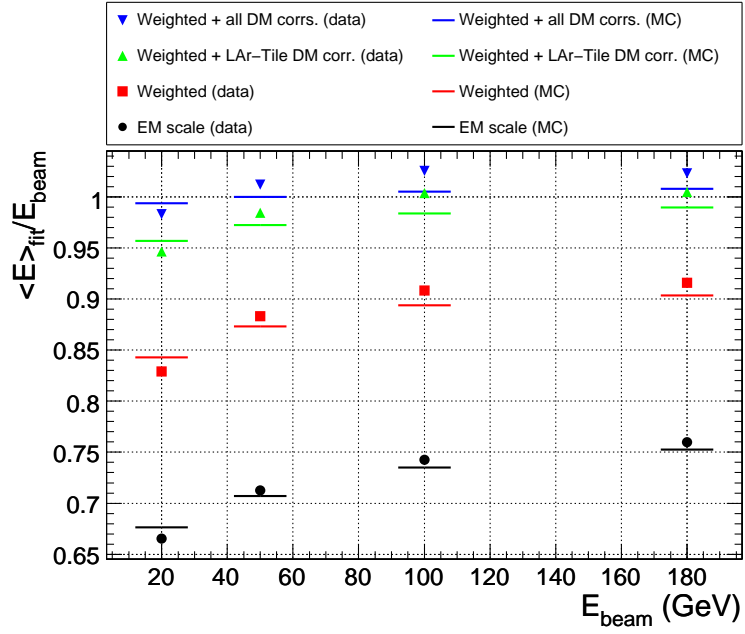}}
    \subfigure[]{\label{fig:resol_data_mc_wholescan}\includegraphics[width=\columnwidth,angle=0]{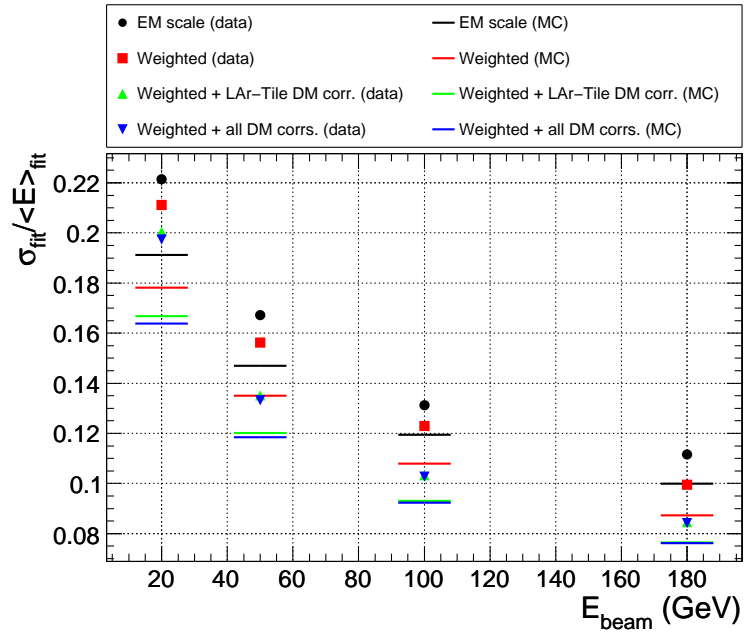}}}
   {\subfigure[]{\label{fig:linearity_data_mc_wholescan}\includegraphics[width=0.7\columnwidth,angle=0]{\imagedir/data_lin.eps}}
    \subfigure[]{\label{fig:resol_data_mc_wholescan}\includegraphics[width=0.7\columnwidth,angle=0]{\imagedir/data_res.eps}}}

  \end{center} \caption{Data and Monte Carlo simulation are compared
  for linearity (a) and relative resolution (b) at all stages of the
  corrections.  The markers show data, while the horizontal lines show
  (a) from bottom to top and (b) from top to bottom: electromagnetic
  scale, compensation weights applied, weight plus LAr--Tile dead
  material correction applied, and all corrections applied. See text
  for details.
  \label{fig:linearity_res_data_mc_wholescan}}
\end{figure}

The performance of the method, as applied to simulation and real beam
test data is shown in Figure
\ref{fig:linearity_res_data_mc_wholescan}. The data at the
electromagnetic scale in this analysis and the one presented in
references~\cite{jenyanote,jenyapaper} are in reasonable
agreement. The largest deviations of about 3\% (2\%) are seen at
20~GeV (180~GeV). At 20~GeV the difference can be explained by the
fact that in that study, the energy in the calorimeters---instead of
using topological clustering---was determined by adding the energies
$E_\textrm{cell}$ of those calorimeter cells having a pseudo-rapidity
within 0.2 of the beam impact point and for which $E_\textrm{cell}$ is
two standard deviations above the expected noise. At 180~GeV, data in
that study were taken with a beam of negatively charged pions, which
does not suffer from proton contamination. In addition, for all beam
energies data were taken in an earlier run period with a different
material configuration upstream of the calorimeters.

The linearity and relative resolution are extracted at all beam
energies for both data and ``mixed signal'' Monte Carlo samples. As in
section \ref{sec:resandlin} the reconstructed energy distribution is
fitted with a Gaussian distribution in the interval ($\mu$ -
2$\sigma$, $\mu$ + 2$\sigma$), where $\mu$ and $\sigma$ are the mean
value and the standard deviation, respectively. Data (simulation) are
shown with markers (horizontal lines) at the electromagnetic scale,
with compensation weights applied, with the dead material correction
for energy lost in dead material between the calorimeters applied, and
lastly at the final calibrated stage, including all dead material
corrections.

After all calibration corrections, the linearity is recovered within
3\% for all beam energies. The discrepancies between data and Monte
Carlo are inherited from the reconstructed energy at the
electromagnetic scale and they are not considerably changed when the
calibration is applied. The relative resolution in data is improved by
about 11\% at low energy (20~GeV) and about 25\% at high energy
(180~GeV) when moving from the electromagnetic scale to the fully
corrected energy scale. A similar relative improvement is obtained in
the Monte Carlo simulation: 14\% at low energy and 24\% at high
energy. The relative resolution is, however, smaller in Monte Carlo
simulation than in data: the discrepancies, at each correction stage,
vary between 10\% and 21\% depending on the energy.  The relative
resolution is smaller in Monte Carlo simulation than in data already
at the electromagnetic scale, by about 10--16\%, depending on beam
energy.  The discrepancies in the shape of the total energy
distribution are more pronounced at lower energies and they are
already present at the electromagnetic scale.

The ratio of data to Monte Carlo simulation is unchanged within 1\%
(4\%) for linearity (resolution) after the corrections are
applied. For linearity such changes are of the same order of magnitude
as the discrepancies between data and Monte Carlo simulation at the
electromagnetic scale: the agreement between data and Monte Carlo
simulation is the same for all correction stages. This means that the
Monte Carlo simulation is able to predict the corrections that should
be applied on the data.

The ability of the Monte Carlo simulation to reproduce the data at the
electromagnetic scale (i.e., before any correction) seems to be the
most critical limiting factor. For the relative resolution the changes
are small, if compared with the discrepancies at the electromagnetic
scale: the discrepancies do not get worse when the corrections are
applied to the data. From preliminary studies a newer Geant4 version
(4.9) is able to provide a better, but still not good, description of
the resolution in the data.

\section{Conclusions}
\label{conclusions}
An energy calibration technique was developed to deal in a coherent
manner with both compensating the hadron response and correcting for
the most significant dead material losses in a segmented
calorimeter. The technique is based on the sensitivity of the
correlation between the deposited energies in the different
calorimeter layers to hadronic and electromagnetic deposits.

The calibration technique was successfully applied to the energy
reconstruction of pions impinging on a subset of the central ATLAS
calorimeters during the ATLAS combined beam test in 2004.  When taking
into account the beam composition of pions and protons, linearity is
recovered within 3\% and relative resolution is improved by between
11\% and 25\%. Consistency with the expectation from Monte Carlo
simulation studies is good for both the linearity and the percentage
improvement in relative resolution. The absolute value of the relative
resolution (after all corrections) is larger in data than Monte Carlo
simulation by 10\% to 21\%.

The discrepancies between data and Monte Carlo simulation are
inherited from the reconstructed energy at the electromagnetic scale
and they are not considerably altered when applying the
calibration. Additional improvement in the data description by Monte
Carlo simulation can help to fulfill the expected absolute value for
the relative resolution.

\section{Acknowledgments}
A very important ingredient of the 2004 ATLAS Combined Beam Test has
been the mechanics of the two calorimeters' support and movement. We
would like to thank Danilo Giugni, Simone Coelli, and Giampiero Braga
from INFN Milano for the design, overview of the production, and
testing of the LAr calorimeter support table. We wish to thank Claude
Ferrari, Pierre Gimenez, Yves Bonnet, Denis Gacon, and Alain Pinget of
CERN EN/MEF group for the continuous mechanical support provided in
the CERN SPS North Area during the installation of the setup and the
data taking.

We are grateful to the staff of the SPS for the excellent beam
conditions and assistance provided during our tests.

We would like to thank Susan Leech O'Neale for proofreading this
paper.

We acknowledge the support of ANPCyT, Argentina; Yerevan Physics
Institute, Armenia; ARC and DEST, Australia; Bundesministerium f\"r
Wissenschaft und Forschung, Austria; National Academy of Sciences of
Azerbaijan; State Committee on Science \& Technologies of the Republic
of Belarus; CNPq and FINEP, Brazil; NSERC, NRC, and CFI, Canada; CERN;
NSFC, China; Ministry of Education, Youth and Sports of the Czech
Republic, Ministry of Industry and Trade of the Czech Republic, and
Committee for Collaboration of the Czech Republic with CERN; Danish
Natural Science Research Council; European Commission, through the
ARTEMIS Research Training Network; IN2P3-CNRS and Dapnia-CEA, France;
Georgian Academy of Sciences; BMBF, HGF, DFG and MPG, Germany;
Ministry of Education and Religion, through the EPEAEK program
PYTHAGORAS II and GSRT, Greece; ISF, MINERVA, GIF, DIP, and Benoziyo
Center, Israel; INFN, Italy; MEXT, Japan; CNRST, Morocco; FOM and NWO,
Netherlands; The Research Council of Norway; Ministry of Science and
Higher Education, Poland; GRICES and FCT, Portugal; Ministry of
Education and Research, Romania; Ministry of Education and Science of
the Russian Federation, Russian Federal Agency of Science and
Innovations, and Russian Federal Agency of Atomic Energy; JINR;
Ministry of Science, Serbia; Department of International Science and
Technology Cooperation, Ministry of Education of the Slovak Republic;
Slovenian Research Agency, Ministry of Higher Education, Science and
Technology, Slovenia; Ministerio de Educaci´on y Ciencia, Spain; The
Swedish Research Council, The Knut and Alice Wallenberg Foundation,
Sweden; State Secretariat for Education and Science, Swiss National
Science Foundation, and Cantons of Bern and Geneva, Switzerland;
National Science Council, Taiwan; TAEK, Turkey; The Science and
Technology Facilities Council and The Leverhulme Trust, United
Kingdom; DOE and NSF, United States of America.

\bibliographystyle{JHEP}
\bibliography{jinstpaper}

\end{document}